\documentclass[twocolumn,amsmath,amssymb,superscriptaddress,aps,prl]{revtex4-1}
\usepackage{mathtools}
\usepackage{graphicx}
\usepackage{braket}
\usepackage{bbold}
\usepackage{tabularx}
\usepackage[dvipsnames]{xcolor}
\begin{document}
%------------------------------------------------------------------------------
\title{
Accurate optical spectra through time-dependent density functional theory based 
on screening-dependent hybrid functionals}

\author{Alexey Tal}
\email{alexey.tal@epfl.ch, alyxthal@gmail.com}
\affiliation{Chaire de Simulation \`a l'Echelle Atomique (CSEA), Ecole
Polytechnique F\'ed\'erale de Lausanne (EPFL), CH-1015 Lausanne, Switzerland}

\author{Peitao Liu}
\affiliation{University of Vienna, Faculty of Physics and Center for
Computational Materials Science, Sensengasse 8, A-1090 Vienna, Austria}

\author{Georg Kresse}
\affiliation{University of Vienna, Faculty of Physics and Center for
Computational Materials Science, Sensengasse 8, A-1090 Vienna, Austria}

\author{Alfredo Pasquarello}
\affiliation{Chaire de Simulation \`a l'Echelle Atomique (CSEA), Ecole
Polytechnique F\'ed\'erale de Lausanne (EPFL), CH-1015 Lausanne, Switzerland}
%------------------------------------------------------------------------------
\begin{abstract}
%------------------------------------------------------------------------------
We investigate optical absorption spectra obtained through time-dependent
density functional theory (TD-DFT) based on nonempirical hybrid functionals that
are designed to correctly reproduce the dielectric function. The comparison
with state-of-the-art $GW$ calculations followed by the solution of the
Bethe-Sapeter equation (BSE-$GW$) shows close agreement for both the
transition energies and the main features of the spectra.  We confront TD-DFT
with BSE-$GW$ by focusing on the model dielectric function and the local
exchange-correlation kernel. The present TD-DFT approach achieves the accuracy
of BSE-$GW$  at a fraction of the computational cost.
%------------------------------------------------------------------------------
\end{abstract}
%------------------------------------------------------------------------------

%------------------------------------------------------------------------------
\maketitle
%------------------------------------------------------------------------------
Semi-local density functionals are notoriously unsuitable
for describing band gaps in semiconductors due to the lack of the derivative
discontinuity \cite{Perdew1983,Sham1983}.  Thus, most of the \textit{ab inito}
methods for optical absorption calculations based on density functional theory 
(DFT) have to address two important problems. First, it is necessary to correct 
the
band gap. Second, the interaction between electrons and holes has to be taken
into account.  The state-of-the-art approach to improve the band gap is the $GW$
approximation~\cite{Hedin1965,Hedin1970,Hybertsen1985,Aryasetiawan1998}, whereas
the electron-hole interaction can be included by solving the Bethe-Salpeter
equation (BSE)
\cite{Onida2002,Albrecht1998,Benedict1998,Rohlfing1998a,Hanke1980}.  The
combined BSE-$GW$ approach has been shown to give very accurate results compared
to experiment, but the main drawback is the scaling, which makes it
computationally very challenging for large size systems. 

Gross developed an alternative approach based on the time-dependent electron 
density, typically referred to as time-dependent density
functional theory~(TD-DFT)~\cite{Onida2002,Gross1985,Runge1984}. In this
approach, the time-dependent Kohn-Sham equations include a time-dependent
exchange-correlation (xc) potential $v_\textrm{xc}$ and its variation with the
time-dependent density, also known as the exchange-correlation kernel 
$f_\textrm{xc}$. The exact $v_\textrm{xc}$ and $f_\textrm{xc}$ are unknown, but 
several approximations have been introduced. Local approximations to 
$f_\textrm{xc}$ lack the correct long wavelength limit, 
$f_\textrm{xc}(q\rightarrow 0) \propto1/q^2$, responsible for the correct 
description of the electron-hole interaction. Therefore, local approximations 
are
unable to capture excitonic effects, but can perform well for metallic
systems \cite{Casida1995,Vasiliev1999,Rubio1996,Gavrilenko1996}.  The correct 
asymptotic behavior is recovered in the so-called ``nanoquanta''
kernel~\cite{Sottile2003,Marini2003,Adragna2003} derived from the BSE to
capture excitonic effects. Hence this approach produces accurate spectra for 
solids, but remains computationally as expensive as solving the BSE. 

Hybrid functional calculations with non-local Fock exchange can be used to
improve band gaps. Moreover, since the long wavelength limit is accounted for,
it can be expected that these functionals could be used for calculating optical
spectra. Various hybrid functionals have been tested in TD-DFT and it has been
shown that a good performance can be achieved in molecules
~\cite{Salzner1997,Laurent2013}. However, a good description in solids requires
the consideration of the screening in the exchange
interaction~\cite{Bruneval2006,Botti2007}. Such a screened interaction was found
to be crucial for the correct description of optical
spectra~\cite{Bruneval2006,Paier2008}.
 
Several different approximations for the screening of the non-local exchange
interaction have been
investigated~\cite{Yang2015b,Refaely-Abramson2015a,Elliott2019,Wing2019,Sun2020}.  
The
results suggest that hybrid functionals yield spectra comparable to BSE-$GW$
provided the adopted fraction of Fock exchange accounts for the screening in the
long range. In particular, Wing~\textit{et al.} obtained good results with
screened range-separated hybrid functionals~\cite{Wing2019}.  However, the 
correct screening in
the short and medium range was not imposed but rather followed from the
empirical setting of the hybrid functional parameters in their TD-DFT approach.
For instance, their choice of taking 25\% of Fock exchange in the short range
does not describe the physically correct behavior of the screening. More
importantly, the range separation parameter was empirically tuned so that the
calculated band gaps matched the $GW$ ones.

Recently, Chen~\textit{et al.}~\cite{Chen2018} developed a nonempirical hybrid 
functional scheme, in which all the parameters are taken from the static 
screening without tuning.
The method showed very accurate electronic structures and band gaps for a large 
number of semiconductors and insulators. The advantage of this approach
is that it accurately accounts for the wave-vector dependent screening: at short 
range the exchange interaction is only weakly screened, whereas in the long 
range it is reduced by the static dielectric constant.

In this work, we investigate the performance of hybrid-functional TD-DFT for 
optical
absorption calculations through the comparison with state-of-the-art BSE-$GW$.
In the TD-DFT scheme, we employ hybrid functionals that have been designed to 
reproduce the correct screening properties through a self-consistent procedure 
\cite{Chen2018}.  We show that this scheme provides absorption spectra with an 
accuracy comparable to that of BSE-$GW$ without tuning parameters.
In particular, we show that equivalent descriptions of the screening in TD-DFT 
and BSE-$GW$ result in very similar absorption spectra.

Following recent work on dielectric-dependent hybrid functionals (DDH)
\cite{Chen2018,Cui2018,Liu2020}, we use in this work the explicit form of the
exchange-correlation potential  given by
\begin{equation}\label{eq:vxc}
\begin{aligned}
V_{\textrm{xc}}\left(\mathbf{r}, \mathbf{r}^{\prime}\right)=
\left[1-\left(1-\epsilon_{\infty}^{-1}\right)\textrm{erf}
\left(\mu\left|\mathbf{r}-\mathbf{r'}\right|\right)
\right]V_{\textrm{x}}^{\textrm{Fock}}\left(\mathbf{r},\mathbf{r'}\right)\\
+\left(1-\epsilon_{\infty}^{-1}\right)
V_{\textrm{x}}^{\textrm{PBE},\textrm{LR}}(\mathbf{r};\mu)
\delta\left(\mathbf{r}-\mathbf{r'}\right)+
V_{\textrm{c}}^{\textrm{PBE}}(\mathbf{r})
\delta\left(\mathbf{r}-\mathbf{r'}\right),
\end{aligned}
\end{equation}
where $\mu$ is the range-separation parameter, $V_{\textrm{x}}^{\textrm{PBE}}$
and $V_{\textrm{c}}^{\textrm{PBE}}$ are the PBE exchange and correlation 
potentials \cite{Perdew1996}.
Here, $V_x^{\textrm{Fock}}$ is the Fock exchange operator given by 
\begin{equation}
V_{\textrm{x}}^{\textrm{Fock}}\left(\mathbf{r,r'}\right)=
-e^{2} \frac{1}{N_\mathbf{k}}\sum_{n\mathbf{k}} 
\frac{\psi^*_{n\mathbf{k}}\left(\mathbf{r}^{\prime}\right) 
\psi_{n\mathbf{k}}(\mathbf{r})}
{\left|\mathbf{r-r'}\right|},
\end{equation}
where $\psi_{n\mathbf{k}}$ are one-electron Bloch states, the sum over 
$\mathbf{k}$ is over $N_\mathbf{k}$ $\mathbf{k}$ points of the Brillouin zone, 
and the sum over $n$ is
over all the occupied bands. In Eq.~\eqref{eq:vxc}, the Fock exchange 
interaction in reciprocal space is thus multiplied by the
function \cite{Liu2020}
\begin{equation}\label{eq:eps}
				c^\text{DDH}_\text{x}\left(\left|\mathbf{q+G}\right|\right) = 
				1-\left(1-\epsilon_{\infty}^{-1}\right)e^{-|\mathbf{q+G}|^2/4\mu^2}.
\end{equation}
In the $GW$ approximation, the Coulomb interaction is screened by the dielectric
function which is a frequency dependent tensor
$\epsilon^{-1}_\mathbf{G,G'}(\mathbf{q},\omega)$ \cite{Aryasetiawan1998}. Thus,
$c^\text{DDH}_\text{x}(|\mathbf{q+G}|)$  in Eq.~\eqref{eq:eps} corresponds to a
model inverse dielectric function
$\epsilon^{-1}_\textrm{model}\left(\left|\mathbf{q+G}\right|\right)$
that neglects the dynamical screening ($\omega \neq 0$) and the off-diagonal 
elements.

\begin{figure}[htpb]
	\centering \includegraphics[width=8.7cm]{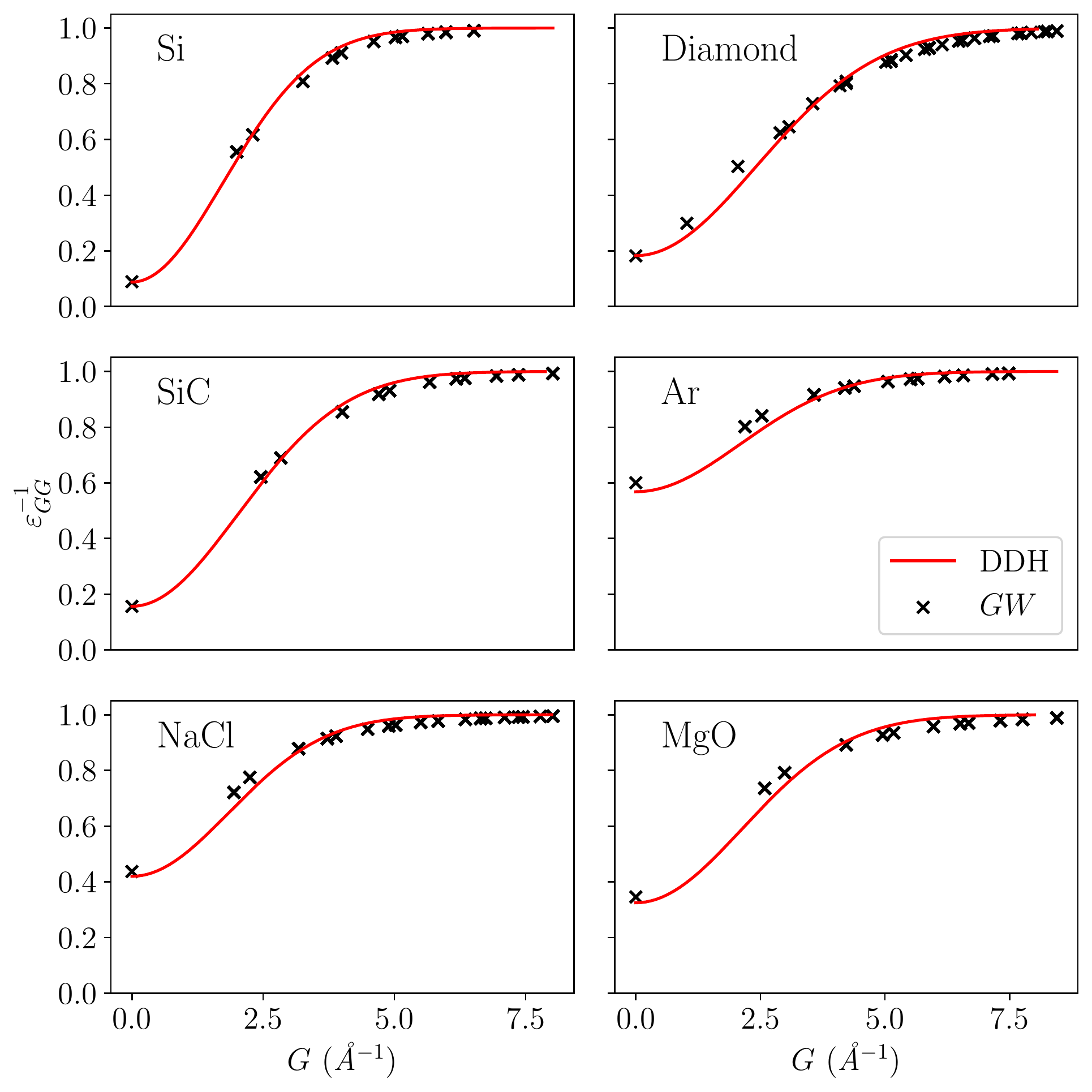} \caption{Inverse 
					dielectric functions vs.\ wave vector at the $\Gamma$
					point for Si, diamond,
	SiC, Ar, NaCl and MgO, as calculated in $GW$ and given by the model in
	Eq.~\eqref{eq:eps} with parameters taken from Ref.\ \cite{Chen2018}.}
	\label{fig:epsilon}
\end{figure}

In the approach of Ref.\ \cite{Chen2018}, the parameters in Eq.\ \eqref{eq:eps} 
are
determined self-consistently. In the long wavelength limit, the interaction is 
set to $1/(\epsilon_\infty q^2)$, where the dielectric constant 
$\epsilon_\infty$ is calculated using the random-phase approximation with vertex 
corrections.  The parameter $\mu$ is obtained by fitting the model to the 
calculated
dielectric function.
In Fig.~\ref{fig:epsilon}, the model dielectric functions associated with the 
DDH are compared to the diagonal elements of the dielectric matrix at the 
$\Gamma$ point at zero frequency.  The dielectric matrix is obtained within 
partially self-consistent $GW$ with vertex corrections [cf.\ Supplemental 
Material (SM)~\cite{Supp}]. In all the cases, we find
the model dielectric function to be in good agreement with the calculated 
dielectric function.

The excitation spectra in both BSE and TD-DFT are obtained by solving an
eigenvalue problem, referred to as the Bethe-Salpeter and Casida equation, 
respectively~\cite{Onida2002,Sander2017}:
\begin{equation}
\label{eq:casida}
\begin{pmatrix}
	\mathbf{A} & \mathbf{B}   \\
	\mathbf{B^*} & \mathbf{A^*} \\
 \end{pmatrix}
\begin{pmatrix}
	\mathbf{X}  \\
	\mathbf{Y}  \\
 \end{pmatrix}
= \Omega
\begin{pmatrix}
	\mathbb{1} & 0  \\
	0 & \mathbb{-1} \\
 \end{pmatrix}
\begin{pmatrix}
	\mathbf{X} \\
	\mathbf{Y} \\
 \end{pmatrix},
\end{equation}
where submatrices $\mathbf{A}$ and $\mathbf{B}$ read
\begin{equation}\label{eq:Aterm}
A_{ai,bj}=(\epsilon_a-\epsilon_i)\delta_{i,j}\delta_{a,b} +\bra{ib}K\ket{aj},
\end{equation}
\begin{equation}
B_{ai,bj}=\bra{ij}K\ket{ab},
\end{equation}
with the indices $i,j$ and $a,b$ referring to occupied and unoccupied states, 
respectively.  The
excitation frequencies of the system are given by $\Omega$. $X$ and $Y$ are
the two-body electron-hole eigenstates in the transition basis
$\psi_a(\mathbf{r})\psi_i^*(\mathbf{r'})$ and
$\psi_i(\mathbf{r})\psi_a^*(\mathbf{r'})$.  Matrix $A$ includes two terms, the
energy of the direct transition from occupied to unoccupied states and the
electron-hole interaction described by the kernel
$K$ \cite{Strinati1984,Rohlfing1998}. Equation \eqref{eq:casida} is
non-Hermitian, which makes it difficult to solve with standard eigenvalue
solvers~\cite{Sander2015,Maggio2016}. A common practice to avoid this difficulty 
is to neglect the coupling
between excitations and de-excitations by setting $B$ to zero. This
approximation is known as the Tamm-Dancoff approximation.

\begin{figure}[t!]
	\includegraphics[width=7.3cm]{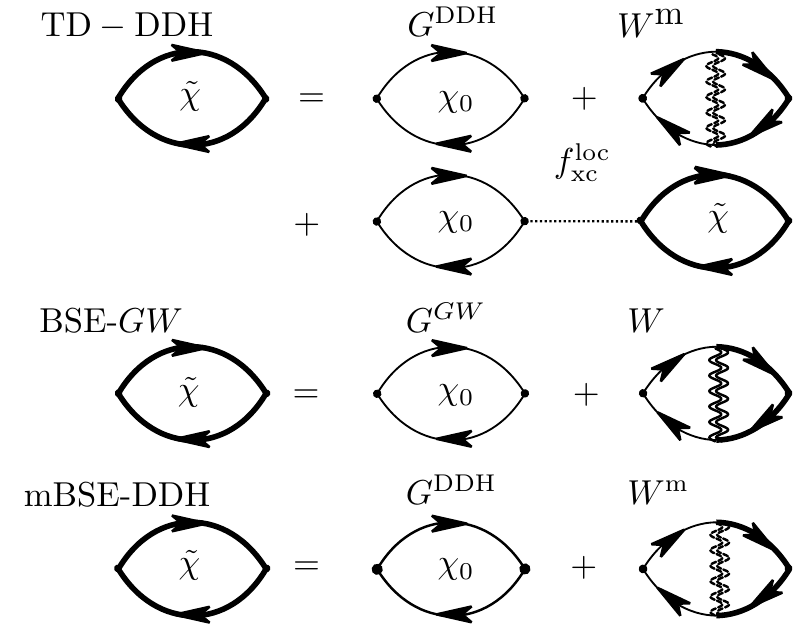}
				\caption{Irreducible polarizabilities $\tilde{\chi}$ in various 
								approximations.
				The reducible polarizability is obtained from
				$\chi=\tilde{\chi}+\tilde{\chi}v\chi$. The wiggly line indicates the 
				screened
				interaction $W$.} \label{fig:diagrams}
\end{figure}

The distinction between BSE-$GW$ and TD-DFT approaches results, on the one hand, 
from the origin of the one-particle eigenfunctions and energies and, on the 
other hand, from the type of the interaction kernel $K$. To make the comparison 
between different methods more transparent,
we provide in Fig.~\ref{fig:diagrams} the Feynman diagrams corresponding to the 
various irreducible polarizabilities $\tilde{\chi}$ discussed in this work. 

In BSE-$GW$, the orbitals and energies are derived from a preceding $GW$
calculation and the kernel consists of a Hartree term $V$ and a screened
exchange term $W$\cite{Rohlfing1998}: \begin{equation}
\bra{ib}K\ket{aj}=2\bra{ib}V\ket{aj}-\bra{ib}W\ket{ja}.
\label{eq:bse}
\end{equation}
The Hartree term describes the bare Coulomb interaction and is the same in all
the approximations considered here. It can be included straightforwardly in a
two-point formulation involving the polarizability $\chi$. The exchange term,
however, requires calculating four-point integrals, which drastically increases
the complexity of the problem.  The screening of the exchange interaction is
determined by the frequency-dependent dielectric function $\epsilon$ obtained
from $GW$ and is represented by a vertical wiggly line in the diagrams. However,
as shown in Refs.~\cite{Rohlfing2000a,Bechstedt1997,Marini2003c}, the
dynamical effects can often be neglected in BSE calculations.

In the TD-DFT approach, the electron energies and wave functions are obtained
from a semilocal or hybrid-functional calculation.  The interaction kernel 
consists of three terms, a Hartree and a screened exchange term $W^{\rm m}$ like 
in the BSE, and an additional local exchange-correlation interaction
$f^\textrm{loc}_{\textrm{xc}}$\cite{Casida2012}:
\begin{equation}
				\bra{ib}K\ket{aj}=2\bra{ib}V\ket{aj}-\bra{ib}W^{\rm
				m}\ket{ja}+\bra{ib}f^\textrm{loc}_\textrm{xc}\ket{aj}.
\label{eq:tddft}
\end{equation}
The screening of the exchange interaction in TD-DFT is described by a constant
through $W^{\rm m}=c_{\rm x}V(|\mathbf{q+G}|)$ or by a function through $W^{\rm
m}=c_\text{x}(\left|\mathbf{q+G}\right|)V(|\mathbf{q+G}|)$,
depending on the exchange-correlation functional.  In particular, $c_\text{x}=0$ 
for semilocal DFT functionals.
In the case of DDH, the exchange interaction is screened by the model inverse 
dielectric function
$c_\text{x} = c^\text{DDH}_\text{x}(\left|\mathbf{q+G}\right|)$ given in Eq.\ 
\eqref{eq:eps} and  the
local exchange-correlation interaction $f^\textrm{loc}_{\textrm{xc}}$ is derived 
from the
local part of the exchange-correlation potential \begin{equation}
f_{\textrm{xc}}^\textrm{loc}(\mathbf{r,r'})=
\frac{\delta\left\{V_{c}^{\textrm{PBE}}+\left(1-\epsilon^{-1}_{\infty}\right)
V_{x}^{\textrm{PBE,LR}}\right\}}{\delta\rho(\mathbf{r})}\delta(\mathbf{r-r'}).
\label{eq:fxc}
\end{equation}
In Fig.~\ref{fig:diagrams}$, f^\textrm{loc}_{\textrm{xc}}$ is
represented by a dotted line connecting $\chi_0$ and $\tilde{\chi}$.
In this work, we refer to this version of TD-DFT as TD-DDH.

\begin{figure}[t]
	\centering
	\includegraphics[width=8.6cm]{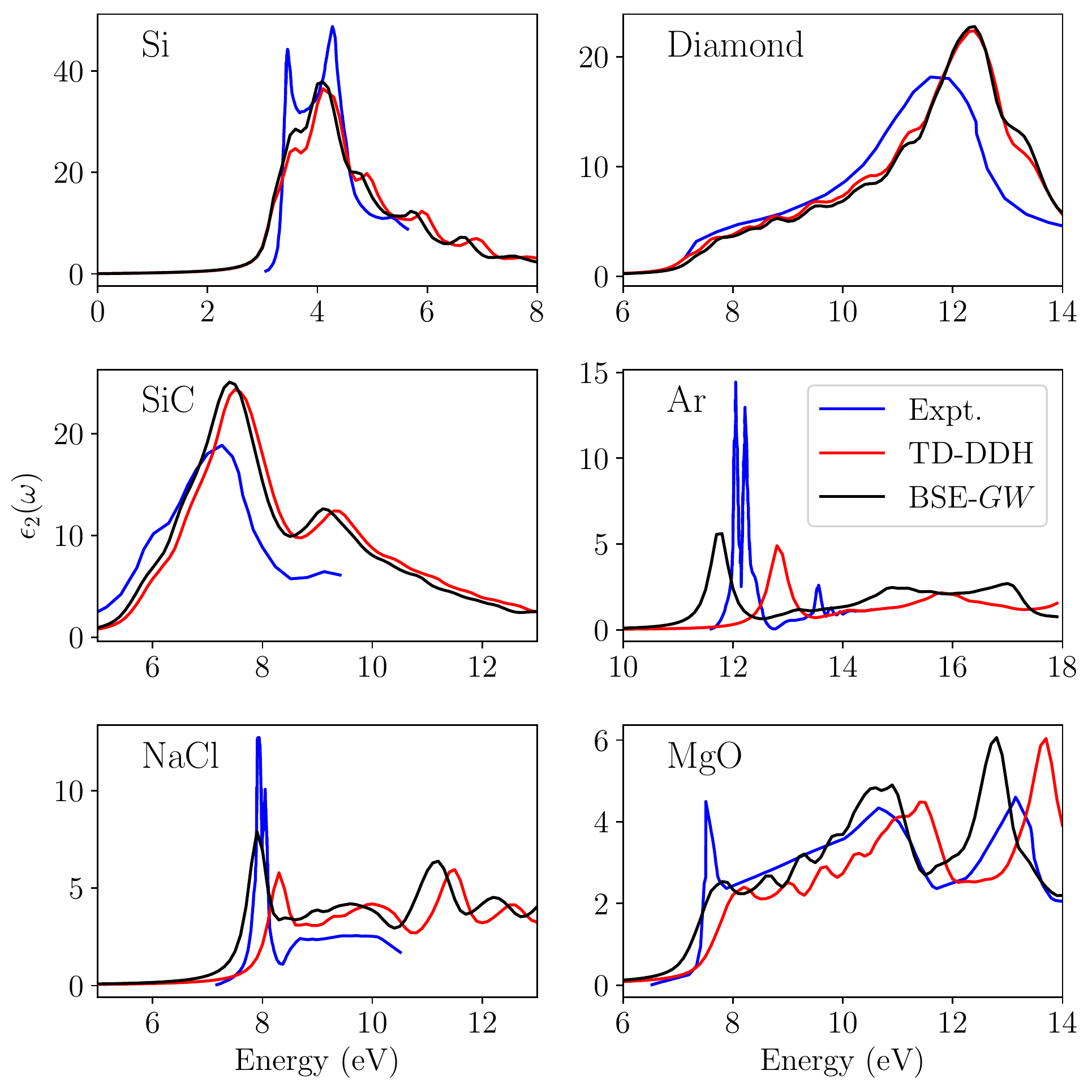}
	\caption{Absorption spectra for Si, diamond, SiC, Ar, NaCl
		and MgO calculated with BSE-$GW$ and TD-DDH.  Experimentally
		measured spectra are taken from \cite{Palik1998} for diamond,
		from \cite{Lautenschlager1987} for Si, from \cite{Logothetidis1996} for SiC, 
		from \cite{Saile1976} for Ar, from \cite{Roessler1968} for NaCl, and from 
		\cite{Bortz1990} for MgO.}	\label{fig:spectra}
\end{figure}

Next, we focus on the comparison between BSE-$GW$ and TD-DDH. In both schemes,
the absorption spectra are obtained from the eigenvalue problem in
Eq.~(\ref{eq:casida}).  In particular, the BSE-$GW$ calculations are based on
partially self-consistent $GW$ using the ``nanoquanta'' vertex corrections
$f_{\rm xc}$ in the polarizability $\tilde \chi$ \cite{Note}.  These two
approaches are tested on a set of materials possessing a wide range of band
gaps. The corresponding spectra are given in Fig.~\ref{fig:spectra}.  Our
calculations show that both approaches agree well with experiment and that
TD-DDH reproduces all the spectral features with the correct oscillator
strengths.  In the case of diamond, Si and SiC, the spectra are nearly on top of 
each
other.  For Ar, NaCl, and MgO, the relative positions of the main features in
the spectra are found to be shifted slightly. In principle, this shift can
result from differences in the band structure and in the screening.  Our
analysis indicates that the dominant effect is due to the energy transition
terms in Eq.\ \eqref{eq:Aterm}.  From Table~\ref{tab:gaps}, we notice that the
calculated band gaps differ by less than 0.15 eV for Si, SiC, and diamond, but 
that
the disagreement is more substantial for NaCl, Ar, and MgO.  Overall, when
compared to experimental values corrected for the coupling to phonons, DDH and
$GW$ band gaps show mean average errors of 0.11 and 0.22~eV, respectively.
These errors are consistent with the current accuracy of {\it ab initio} methods
\cite{Shishkin2007,Chen2015}, indicating that the agreement with experiment
should be considered excellent for both schemes.  As far as the screening is
concerned, we show below that the small discrepancies observed in
Fig.~\ref{fig:epsilon} hardly change the spectra.

% comparison with other td-hyb
The performance of TD-DDH can also be assessed through a comparison with TD-PBE0, an
approach commonly used for the calculations of spectra \cite{Voros2009,Yang2015b}.
TD-PBE0 is based on a global hybrid functional where 25\% of Fock exchange is
used uniformly and the exchange interaction in the calculation of the spectra is
screened by $\varepsilon_\infty$.  As shown in the SM \cite{Supp}, the
accuracy of TD-DDH is significantly better. Moreover, TD-PBE0 does not reproduce
the dielectric screening over the full range, which obscures the understanding
of the underlying physics. 

\begin{table}[] \caption {Band gaps (in eV) obtained with the semilocal PBE
				functional \cite{Perdew1996}, DDH, and $GW$. The experimental values are
augmented by theoretical corrections resulting from the coupling to phonons.}
	\begin{tabularx}{\columnwidth}{XXXXXXXX}
	\hline\hline
							& Si   & SiC  & Diamond &   & NaCl & Ar    & MgO  \\\hline
					PBE & 0.75 & 1.35 & 4.14 && 5.21 & 8.70  & 4.77 \\ DDH & 1.31 & 2.50 & 
					5.69 && 9.13 & 14.60 & 8.41 \\
					$GW$ & 1.41 & 2.55 & 5.85 && 8.86 & 13.75 & 8.12 \\ Expt. & 
					1.23\footnote{Ref.~\cite{Bludau1974}, with a correction of 0.06 eV 
	from Ref.~\cite{Cardona2005}.}&
	 2.53\footnote{Ref.~\cite{Humphreys1981}, with a correction of 0.11 eV from 
	 Ref.~\cite{Monserrat2014}.} &
	 5.85\footnote{Ref.~\cite{Clark1964}, with a correction of 0.37 eV from 
	 Ref.~\cite{Cardona2005}.} &&
		9.14\footnote{Ref.~\cite{Roessler1968}, with a correction of 0.17 eV from 
		Ref.~\cite{Lambrecht2017}.} &
		14.33\footnote{Ref.~\cite{Baldini1962}, with a correction of 0.03 eV
	calculated using the method described in 
	Refs.~\cite{Zacharias2016,Karsai2018}.} &
		8.36\footnote{Ref.~\cite{Hinuma2014}, with a correction of 0.53 eV from 
		Ref.~\cite{Nery2018}.} \\ \hline\hline
	\end{tabularx}
	\label{tab:gaps}
\end{table}

\begin{figure}[htpb]
	\centering
	\includegraphics[width=8cm]{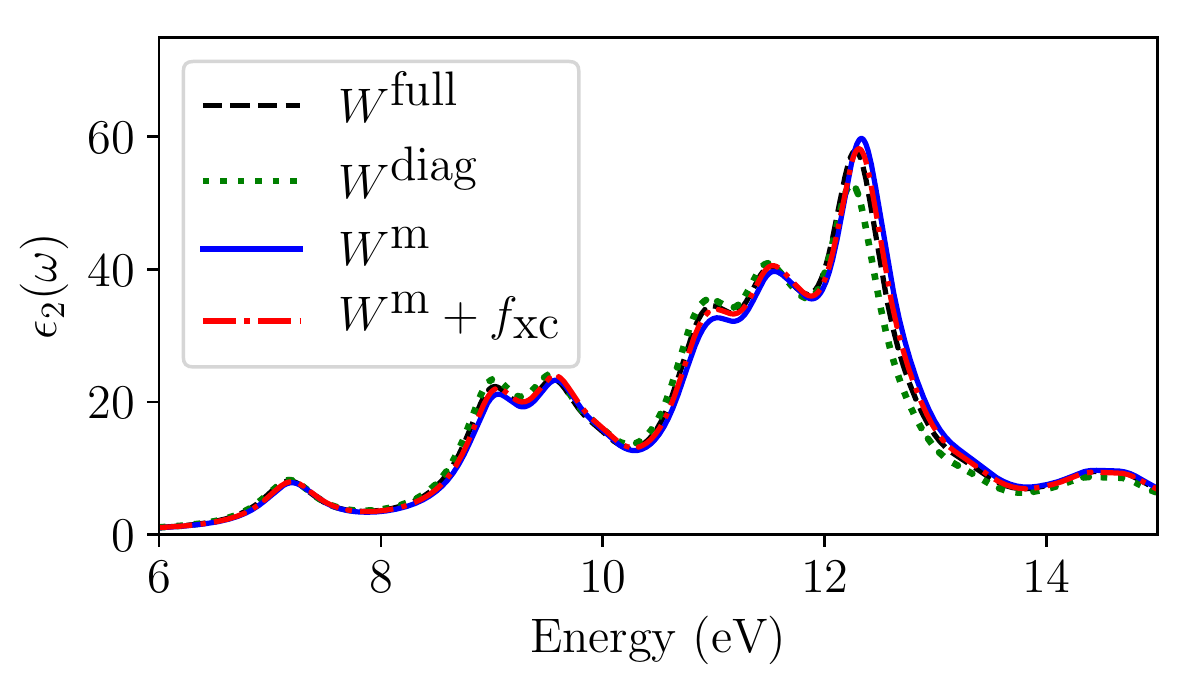}
	\caption{Optical absorption spectrum of diamond calculated with BSE and full
		$\epsilon^{-1}$ from $GW$ ($W^\mathrm{full}$), BSE and diagonal
		$\epsilon^{-1}$ ($W^\mathrm{diag}$), BSE and model
		$\epsilon^{-1}_\textrm{model}$ ($W^\textrm{m}$),  TD-DDH and
		model $\epsilon^{-1}_\textrm{model}$ with $f^\textrm{loc}_\textrm{xc}$ 
		($W^\textrm{m}+f^\textrm{loc}_\textrm{xc}$).
		All spectra are based on energies and wave functions from a
		$G_0W_0$ calculation. A $6\times6\times6$ {\bf k}-point grid is used.}
		\label{fig:fxc}
\end{figure}

% Fig. 4 - screening issues
To compare the screening in BSE-$GW$ and TD-DDH, we show in Fig.\ \ref{fig:fxc} 
the absorption spectra of diamond calculated in various approximations using the 
same energies and wave functions, which are taken from a $G_0W_0$ calculation.  
We start our analysis from a BSE calculation in which the full
static inverse dielectric matrix is used ($W^{\rm full}$). In particular, we 
show that the off-diagonal elements of this matrix barely have any effect on the 
calculated spectrum ($W^{\rm diag}$), in accordance with Ref. \cite{Sun2020}.  
Next, we replace the inverse dielectric function with the model 
$c^\text{DDH}_\text{x}\left(\left|\mathbf{q+G}\right|\right)$
and find no discernible difference in the spectrum ($W^{\rm m}$).  Notice that 
we here consider isotropic screening and that the extension of this model to
anisotropic materials remains to be investigated.
The present treatment of the screening is equivalent to that of TD-DDH, where 
the local exchange-correlation kernel is neglected, i.e.\ 
$f^\textrm{loc}_\textrm{xc}=0$,
also referred to as model BSE (mBSE) \cite{Bokdam2016,Sun2020}.
To restore the full TD-DDH screening, we include the 
$f^\textrm{loc}_\textrm{xc}$ and still obtain essentially the same spectrum 
($W^{\rm m} + f^\textrm{loc}_\textrm{xc}$).
Hence, these results indicate that the model screening in TD-DDH gives an 
accurate description of the screening in $W= \epsilon^{-1} V$ and that the 
effect of $f^\textrm{loc}_\textrm{xc}$ is negligible for extended systems.  
Considering
this, we can say that for extended systems TD-DDH is {\it de facto} equivalent 
to mBSE.

% Numerical complexity
The numerical complexity of Eq.~(\ref{eq:casida}) is the same in BSE and TD-DDH.
However, the preceding $GW$ calculations required in BSE-$GW$ involve a high
computational cost, which scales like $N^4$ in the number of electrons $N$ in 
most
$GW$ implementations instead of like $N^3$ in TD-DFT.  Additionally, in the
calculation of the Green's function  in $GW$, the convergence with respect to
the number of unoccupied states and  the number of frequency points has to be
controlled carefully, which significantly increases the complexity of the
calculations. Note that the static dielectric constant only needs to be
determined at the $\Gamma$ point of the Brillouin zone and that it converges
quickly with respect to the number of included orbitals.  Furthermore, the
hybrid-functional approach only requires a model static dielectric function, for
which the static limit can be obtained rather  efficiently \cite{Cui2018}. Thus,
the hybrid functional approach opens the way to more efficient numerical schemes
that can circumvent the calculation of the full dielectric matrix. 

% Conclusion
In conclusion, we have shown that time-dependent calculations using the 
parameter-free DDH functional yield optical absorption spectra with an
accuracy comparable to BSE-$GW$. The success of this approach originates from 
the use of a model dielectric function that gives a physically motivated 
description of the screened exchange interaction over the full spatial range.  
Notably, the computational complexity of the method is drastically reduced 
compared to BSE-$GW$, as it eliminates the need for preceding $GW$ calculations.  
This will allow one to consider larger and more complex systems than hitherto 
possible. 

The structures and the input files used for the calcula-
tions are freely available on the Materials Cloud platform,
see Ref. \cite{Cloud}.

Support from the Swiss National Science foundation is acknowledged under Grant
No.\ 200020-172524.  The calculations have been performed at the Swiss National
Supercomputing Centre (CSCS) (grant under project ID s879) and at SCITAS-EPFL.

%------------------------------------------------------------------------------
\bibliography{article}

\begin{thebibliography}{67}
\bibitem [{Not()}]{Note}%
  \BibitemOpen
  \href@noop {} {}\bibinfo {note} {{Fully self-consistent $GW$ as in QS$GW$
  does not lead to appreciable differences in the spectra \cite{Supp}
  }}\BibitemShut {NoStop}%
\end{thebibliography}
%------------------------------------------------------------------------------
\end{document}